\documentclass{article}

\usepackage{PRIMEarxiv}

\usepackage[utf8]{inputenc} 
\usepackage[T1]{fontenc}    
\usepackage{hyperref}       
\usepackage{url}            
\usepackage{booktabs}       
\usepackage{amsfonts}       
\usepackage{nicefrac}       
\usepackage{microtype}      
\usepackage{lipsum}
\usepackage{fancyhdr}       
\usepackage{graphicx}       
\graphicspath{{media/}}     
\usepackage{cite}
\usepackage{amsmath,amssymb,amsfonts}
\usepackage{algorithmic}
\usepackage{graphicx}
\usepackage{textcomp}
\usepackage{float}
\usepackage{subfloat}
\usepackage{subfig}
\usepackage{caption}
\usepackage{color,soul}

\DeclareUnicodeCharacter{0301}{\'{e}}

\def\BibTeX{{\rm B\kern-.05em{\sc i\kern-.025em b}\kern-.08em
    T\kern-.1667em\lower.7ex\hbox{E}\kern-.125emX}}
    
\title{Multi-path fading and interference mitigation with Reconfigurable Intelligent Surfaces}

\author{
  Jean Baptiste Gros, Geoffroy Lerosey \\
  Greenerwave, 75005 Paris, France \\
  \And
  Fabrice Lemoult\\
  ESPCI Paris, PSL University, CNRS \\ 
  The Institut Langevin, 75005 Paris, France \\
   \And
  Mir Lodro, Steve Greedy, Gabriele Gradoni \\
  Radio Frequency Laboratory \\ 
  George Green Institute for Electromagnetic Research-GGIEMR\\  
  University of Nottingham, NG72RD,  United Kingdom \\
}

\begin{document}
\maketitle

\begin{abstract}
We exploit multi-path fading propagation to improve both the signal-to-interference-plus-noise-ratio and the stability of wireless communications within electromagnetic environments that support rich multipath propagation.  
Quasi-passive propagation control with multiple binary reconfigurable intelligent surfaces is adopted to control the stationary waves supported by a metallic cavity hosting a software-defined radio link. 
Results are demonstrated in terms of the error vector magnitude minimization of a quadrature phase-shift modulation scheme under no-line-of-sight conditions.
It is found that the magnitude of fluctuation of received symbols is reduced to a stable constellation by increasing the number of individual surfaces, or elements, thus demonstrating channel hardening. 
By using a second software-defined radio device as a jammer, we demonstrate the ability of the RIS to mitigate the co-channel interference by channel hardening.
Results are of particular interest in smart radio environments for mobile network architectures beyond 5G.
\end{abstract}

\begin{figure}
\centering
\includegraphics[width=0.5\columnwidth]{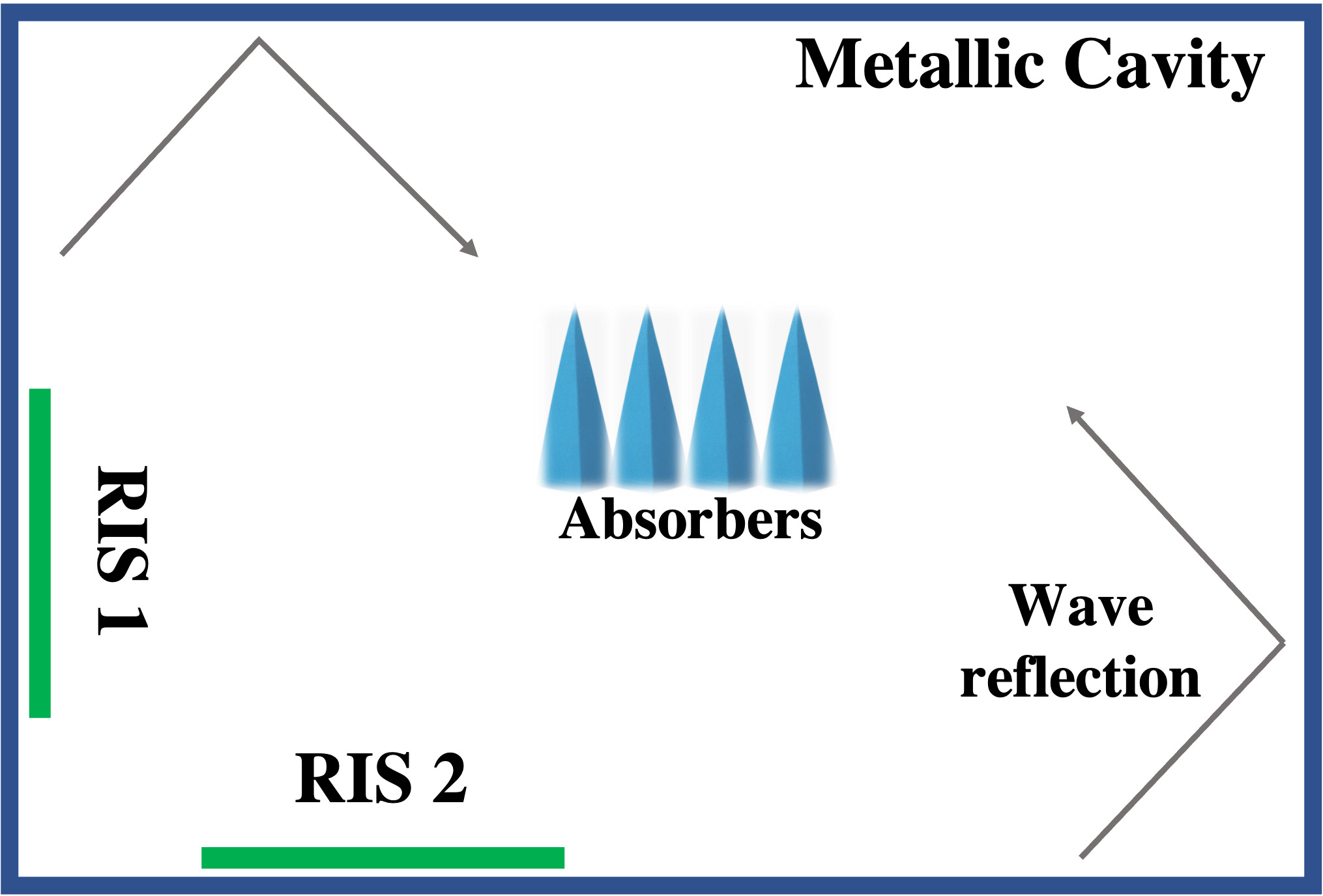}
\caption{Metallic cavity equipped with wall mounted binary RISs. The objects and absorbers inside the cavity create non-uniform wave field distribution by reflection and scattering. \label{fig:fig0}}
\end{figure}

Reconfigurable intelligent surface (RIS) technology has opened up a new approach to achieving efficient, i.e., high data-rate and high capacity, wireless communications \cite{DiRenzoApr2020,Zhang2019}. 
Quasi-passive planar structures equipped with tunable reflecting 
metallic patches were first designed and engineered in \cite{Kaina2014,Kaina2014a}, to achieve electromagnetic energy focusing within wave chaotic environments \cite{DelHougne2016}. 
The concept of smart environments was first introduced in \cite{Subrt1,Subrt2}, where the authors proposed to optimize the usage of various access points inside a building, by controlling the electromagnetic reflectivity of its walls in order to distribute the signal in an optimal way within the building.
Later on, the idea of using electronically reconfigurable surfaces for improved wireless communications was proposed in \cite{Kaina2014,Dupre2015}, where the fields incident on the surface were locally controlled in order to focus them on a given antenna.
Follow up works were then published over the years, for example \cite{X.Tan,Welkie,Hougne}, but the topic remained of little interest to the wireless community, until a series of papers were published in late 2018 and 2019 \cite{Hougne,Liaskos_7,Liaskos_8,Renzo2019,Basar2019}.
From there, the idea of smart environments using electronically reconfigurable surfaces, coined Reconfigurable Intelligent Surfaces (RIS) emerged as a credible major technological advancement for 6G and has attracted an enormous and growing interest in the wireless communications community. Associated to RIS there are numerous research topics ranging from energy efficient wireless communication\cite{8741198} to channel modeling \cite{DiRenzo2020,Garcia2020}, RIS based signal modulation and encoding \cite{Basar2020},  MIMO channel estimation and beamforming \cite{Nadeem2020,He2020,Park2020}, telecommunication  performance evaluation \cite{Badiu2020}, mathematical model  and  optimization method for wavefront shaping with RIS \cite{Wu2020,Zappone2020,Huang2019} and even stochastic analysis approaches \cite{DiRenzo2019}.
Of the numerous works proposed, a very large proportion are concerned with theoretical and mathematical approaches, and very few deal with experimental demonstrations of RIS \cite{StrinatiRISE6G2021}.
However, the technology has been patented as an innovative antenna for use in space communication links \cite{GrosRIS2021,gros2021wave}. 
Other, but limited experimental realisations of meta-surfaces exist 
in the scientific literature and we mention here the spatially modulated meta-surfaces adopted in beam-forming \cite{Minatti2015} and 
large printed dipoles antenna arrays for long range focusing 
\cite{Arun2019}. 
Previous works looked to develop a beam-steering approach based on wave emission, while in this work we propose to control the propagation of waves emitted by a simple antenna. 
Furthermore, while those realisations remain a valid tool for beam-forming in free-space, the RIS adopted in the present study 
possess dual polarisation unit cells controlled 
by an in-house sequential optimisation algorithm \cite{Vellekoop2007,GrosRIS2021,PopovExtender2021} that accounts for the back-reflection from the multi-path propagation environment \cite{Gros2020a}. 
Of significant interest is the fact that  the meta-surface can be optimised globally through embedded electronics, e.g., FPGA platforms, integrated within the unit cell forming the planar meta-surface \cite{Tsilipakos2020}.
Further RIS applications include 
energy harvesting \cite{DelHougne2017}, motion detection \cite{DelHougne2018b}, as well as environment based 
digital processing and computing \cite{DelHougne2018a}.
More recently, the ability of the binary RIS to engineer the 
eigen-space of the environment has been exploited to create 
wave chaos in integrable environments \cite{Gros2020a}
and electronic mode-stirring in static reverberation chambers 
\cite{Gros2020b}.
The electromagnetic (EM) environment that we consider in this study is inherently an overmoded metallic cavity (MC), equipped with two wall-mounted binary RIS elements.
The MC offers a flexible environment for the 
emulation of a real-world multi-path propagation environment \cite{Valenzuela2008,HollowayFading2010,Arslan2019}.
This is depicted in Fig. \ref{fig:fig0}.

For instance, the power delay profile (PDP) of any environment that one could encounter in practice can be reproduced statistically 
inside an MC by tailoring the losses within the metallic cavity sustaining the wireless energy \cite{Mariani_2020} where a two-antenna universal software radio peripheral (USRP) is operated inside the MC to create a digital transmission link and an additional two-antenna USRP is used to create intentional co-channel interference disturbing the legitimate MC link. 
Of additional interest is the experimental work on RIS empowered wireless communications that reports path loss (PL)
as well as scattering parameter studies in anechoic and indoor scenarios. 
It is worth mentioning, for examples, the PL model validation 
in \cite{Tang2020} as well as the transmission performance 
of static metallic reflectors in \cite{Khawaja2020}.
Here, we focus on the study of performance of the digital 
transmission based on parameters monitored by the USRP devices. 
The study fills the gap between proofs of principle achieved in the applied physics community and the system predictions obtained by the wireless communications community. 
This is a fundamental step towards transferring the technology towards practical application in mobile network architectures. 
Notable work that show related studies is present in the literature. 
A recent RIS-based transceiver has been developed, building on 
conventional embedded electronics systems in \cite{Tang2020}. 
Furthermore, modulation of commodity Wi-Fi signals has been used to deliver a wireless digital transmission based on SDR \cite{Zhao2020}. Authors in \cite{tang2019programmable} have demonstrated a programmable metasurface-based transmitter to show comparable BER performance and transmission rates as more conventional approaches.

The scope of this article is to show how RIS can be used to improve the signal-to-interference-plus-noise ratio (SINR) by reducing the effect of inference in, and due to, the environment while increasing the local energy density focusing. 
We then widen this scope by considering intentional, alongsides unintentional and environment driven interference mechanisms. 
Achieved results include an experimental proof of channel hardening in RIS-assisted single-input single-output (SISO) wireless links \cite{Bjornson2020} and that robustness against arbitrary dynamic intense interference can be achieved by deploying multiple RISs.

The article is organised as follows. In Section I we revisit the RIS technology, the software defined radio system, the controlled propagation environment, as well as the practical realisation of the wireless link in Section II. In Section III, we provide a detailed account of the results obtained with and without intentional interference. Finally, in Section IV, we draw some conclusions and future perspective. 

\section{RECONFIGURABLE TECHNOLOGY AND METHODS}
The wireless link is a single-input single output wireless channel 
operating within a metallic cavity with variable losses. 
Following tradition, and as adopted in \cite{Guan2020,Espinosa2020}, the transmitter is hereby referred to as 
Alice, the receiver is referred to as Bob, while the intentional interferer in the external environment is referred to as Eve. 

\subsection{RIS assisted multipath environment}
\begin{figure}
\centering
\includegraphics[width=0.5\columnwidth]{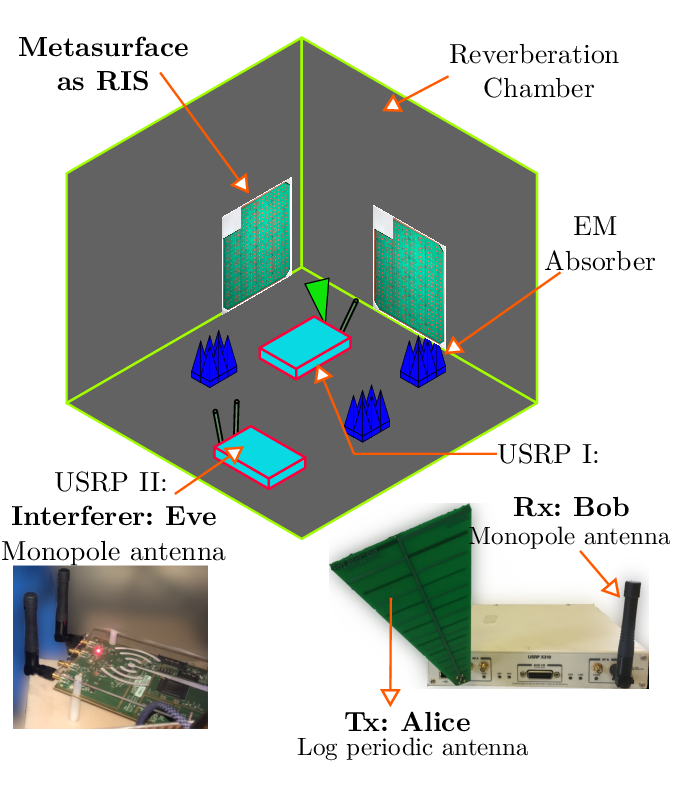}
\caption{Metallic cavity equipped with wall mounted binary RISs. The URSP X310 supporting the Alice-Bob link operates in presence of a second USRP supporting the co-channel interferer Eve. \label{fig:fig1}}
\end{figure}

The EM environment within the MC is presented in Figure \ref{fig:fig1}. 
The MC that we have used, is a large metallic enclosure, i.e., of characteristic  size that is much larger than the wavelength of the excitation (carrier) frequency.
In electromagnetic compatibility (EMC) applications, an MC is often operated under mechanical mode-stirring by moving an irregular metallic structure though a number of  statistically independent positions \cite{Mariani_2020}. 
In this work, we leverage the multipath fading generated by a high density of cavity eigenmodes excited at the carrier frequency within the MC. 
The MC is loaded with EM absorbing material, depicted 
in Fig. \ref{fig:fig1} as blue cones.
A variable amount of absorbing material is typically used to emulate the PDP of real-life outdoor/indoor environments. 

An USRP X310 maintains the wireless digital link between 
two antennas in the same device. 
Two antennas are chosen to achieve a cross-polarized configuration, thus 
minimising the direct coupling within the device, 
i.e., a no line of sight situation is created. 
Also, while a second USRP is present, its operation 
is initially switched off in order to study the 
unperturbed propagation channel. 
This means that Alice and Bob can communicate without 
disturbance from Eve. 
Once switched on, Eve creates a real attack to the 
wireless link by sending random noise 
within the same frequency band supporting the Alice-Bob link. 

\subsection{RIS technology and optimisation}
\begin{figure}
    \centering
	\includegraphics[width=0.5\columnwidth]{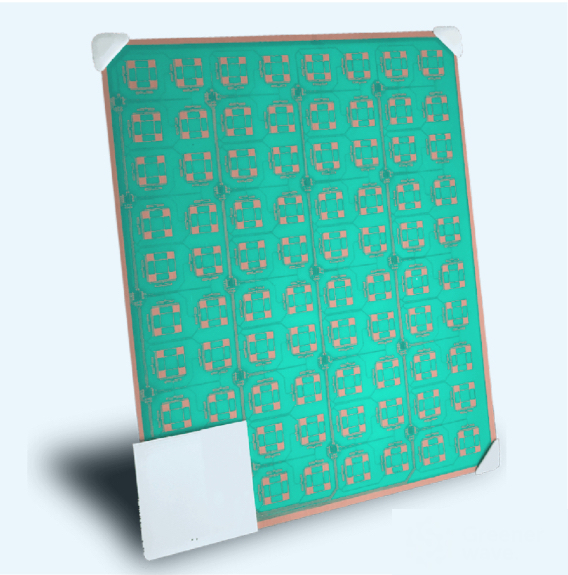}
	\caption{Picture of the $5.2$~GHz RIS used in this experiment. The surface consists of 76 phase-binary pixels. And each pixel can be configured electronically to control independently, via the states of two PIN diodes, both components of the reflected electric field tangential to the pixel plane thus leading to a total number of 152 effective pixels.  \label{fig:RIS1}}
\end{figure}
The RIS is a 6G enabling technology that can configure the wireless propagation environment by using low-cost passive elements. 
Those elements perform independent fine tuning of the phase shifts, as well as precise control of amplitude levels of the reflected waves.  
The collective response of an array of passive elements can be engineered for example to beamform, focus, or cancel the wireless signal propagating in complex EM environments. 
The RIS structure that we consider here is based on a binary unit cell that has been demonstrated in free-space \cite{GrosRIS2021} and closed environments \cite{Dupre2015}.
Most of the theoretical work considers higher phase shift values and fine amplitude level control, however, in practical terms this may introduce a lot of overhead. 
In this work we use a maximum of 2 RISs operating at sub-6GHz frequencies, in the layout represented in Figure~\ref{fig:RIS1}.  
The RIS are connected to a PC via USB that is used to simultaneously provide power to the integrated control board of the RIS and send commands to configure the RIS by controlling the bias voltage of a set of PIN diodes through a set of shift registers.  Each surface consists of 76 phase-binary pixels, which are based on hybridizing two resonances \cite{Kaina2014,Kaina2014a}. Each pixel can be configured electronically to control independently, via the states of two PIN diodes, both components of the reflected electric field tangential to the pixel plane by imposing a 0 or $\pi$ phase shift, thus leading to a total number of 152 effective pixels per RIS. Since the design of the RIS is based on resonant effects, the frequency range over which they are efficient is limited to 1 GHz around 5.2 GHz. The  RIS design shown in Figure~\ref{fig:RIS1} has already featured in several experimental investigative works ranging from wave chaos physics to EMC and wireless communications \cite{Gros2020a,GrosRIS2021,lodro2021reconfigurable} and its working principles are described in the following papers \cite{Kaina2014,Kaina2014a}. 
The RISs are used to improve the communication between Alice and Bob and  to that end, we use an optimisation procedure to find the RIS configurations that will improve the quality of the data link. 
This procedure, described later in the paper, requires the measurement of quantities related to the performance of the data links. Different approaches exist, for instance  a pre-optimisation of the RIS configurations, through channel sounding, via the evaluation of the scattering matrices between Alice and Bob using a vectorial network analyser \cite{PopovExtender2021,Emil_2021a}, that is subsequently replaced by an actual communication set-up.  
However as we are using a SDR based techniques and a USRP to establish the link, we will have direct access to key performance  indicators (KPI) of the communication link such as bit-error-rate (BER) or error vector magnitude (EVM). 
Therefore, we can use these KPIs to simultaneously probe the quality of the communication channel and optimize it by adjusting the RIS configuration.

\subsection{Software Defined Radio Setup}
We used two USRPs, hereafter referred to as USRP I and USRP II, for creating a wireless communication data link and intentional interference respectively. 
A X310 USRP (USRP I) with UBX-160 RF daughterboard is a high performance full-duplex SDR that has been used to sustain a high data-rate 2x2 MIMO communication link. 
Another SDR, again from Ettus Research, a B210 USRP (USRP II) with integrated RF front-end is used to generate the interference signals. 
The USRP II is also a full-duplex SDR that can also maintain 2x2 MIMO link, however at low sampling rates in contrast to the USRP I.  
We connected the Tx antenna (Alice) with channel 1 of USRP I and Rx antenna (Bob) with channel 2 of USRP I, in order to avoid intra-board leakage. 
More precisely, a log periodic antenna was used for Alice and a dipole antenna was used for Bob.
Both the antennas are placed in cross-polarized orientation to avoid direct coupling between them.Therefore, there is no direct link between Alice and Bob, neither via internal SDR cross-talk, nor through direct LOS.

\begin{figure}
    \centering
    \subfloat[]{\label{fig:alice}\includegraphics[width=0.7\columnwidth]{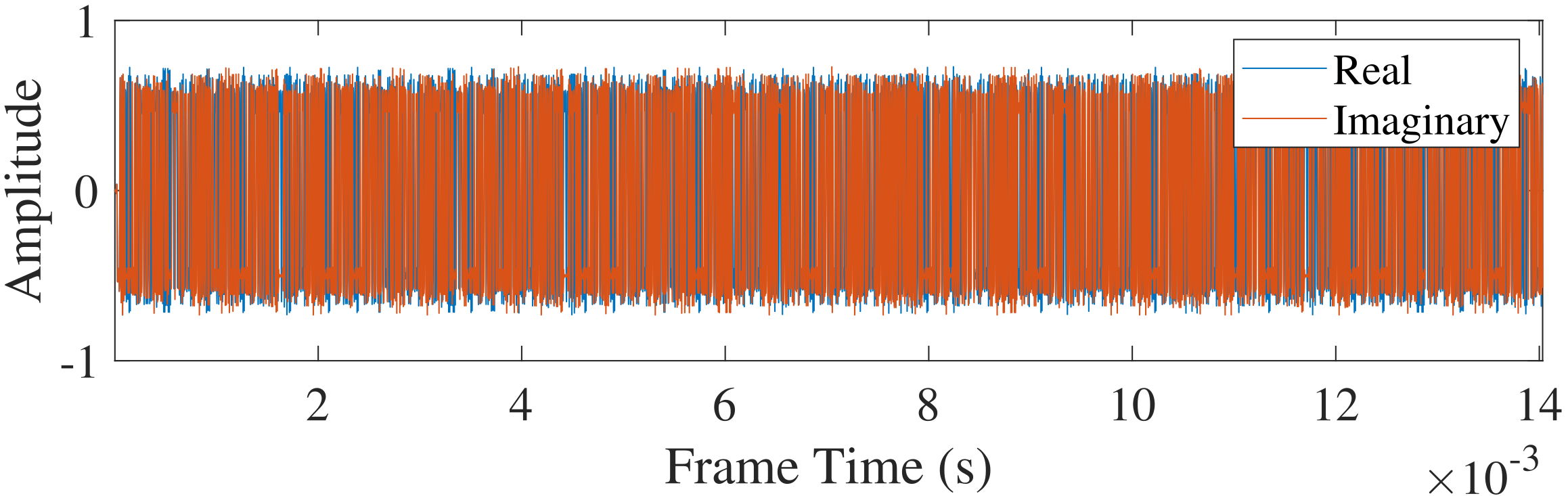}}\\
    \subfloat[]{\label{fig:eve}\includegraphics[width=0.7\columnwidth]{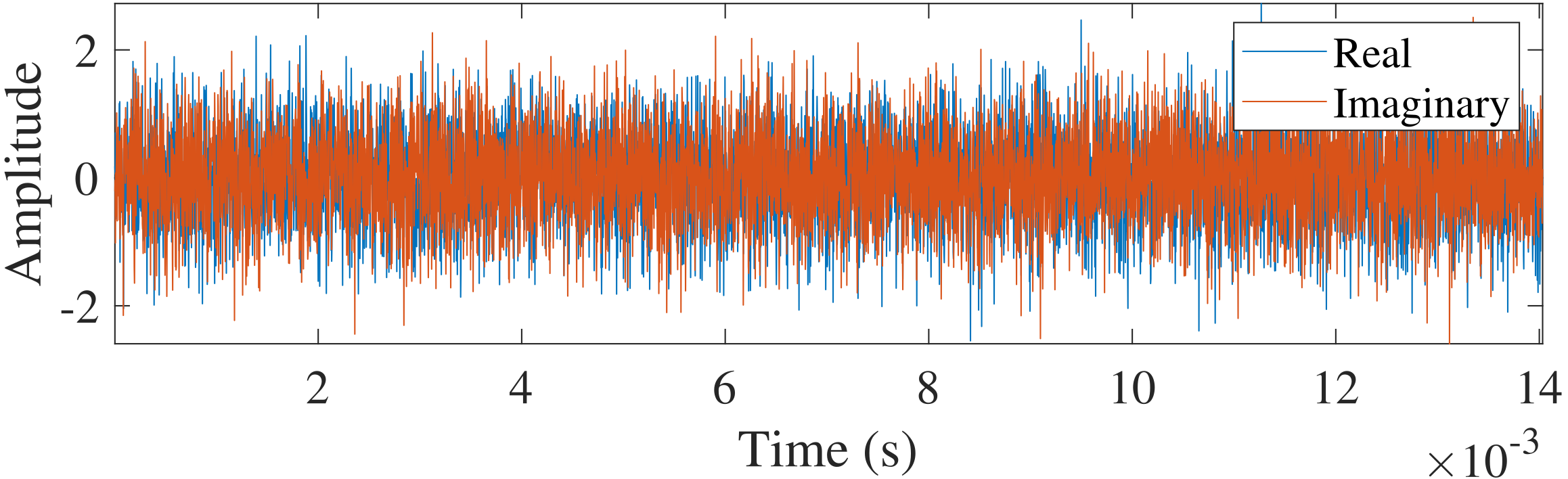}}
    \caption{Alice and EVE waveforms (a) typical QPSK modulated waveform as sent by Alice (b) snapshot of EVE noise waveform.}
    \label{fig:alice_eve_waveform}
\end{figure}

\begin{figure}
    \centering
   \subfloat[]{\includegraphics[width=0.4\columnwidth]{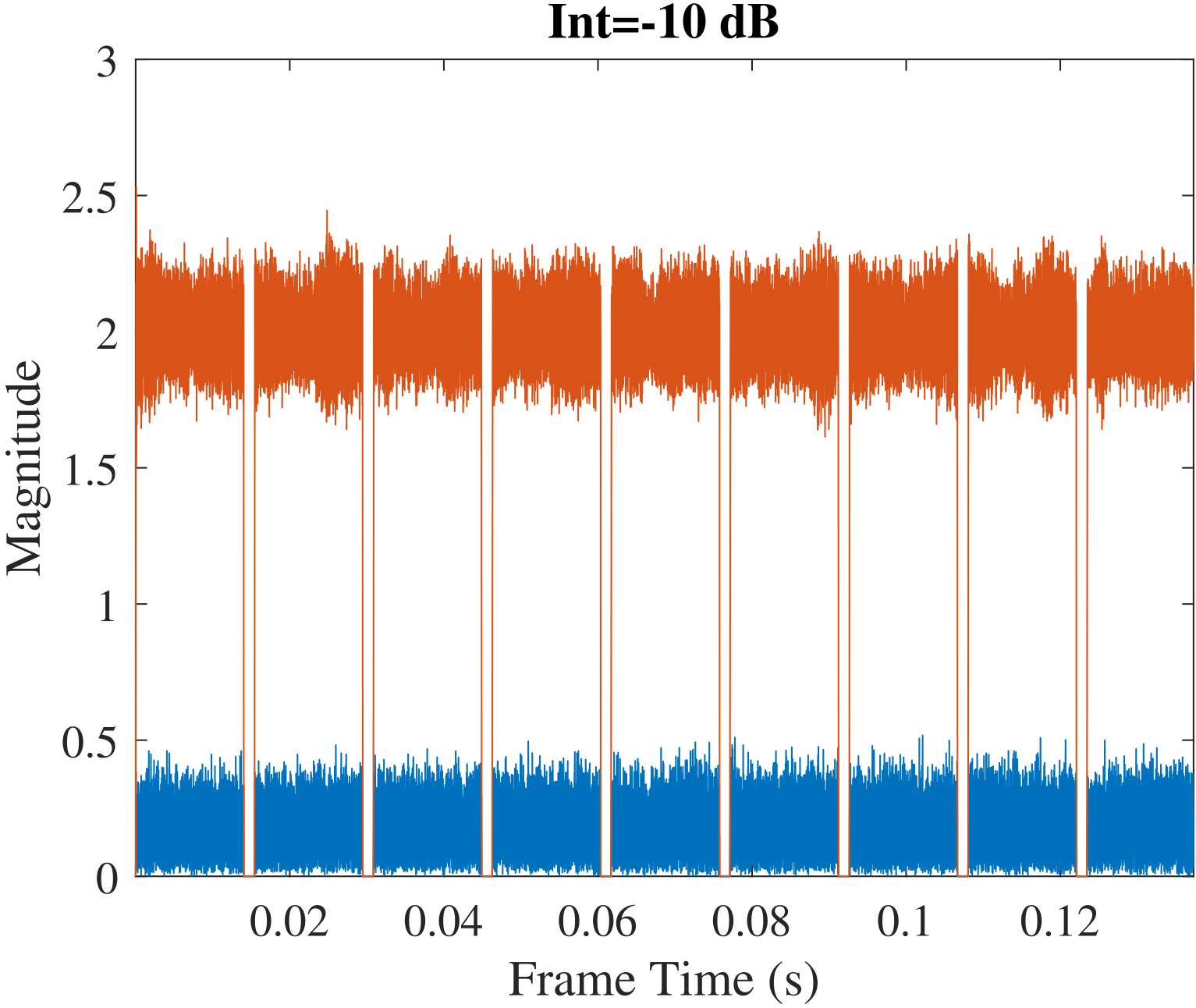}}\\
   \subfloat[]{\includegraphics[width=0.4\columnwidth]{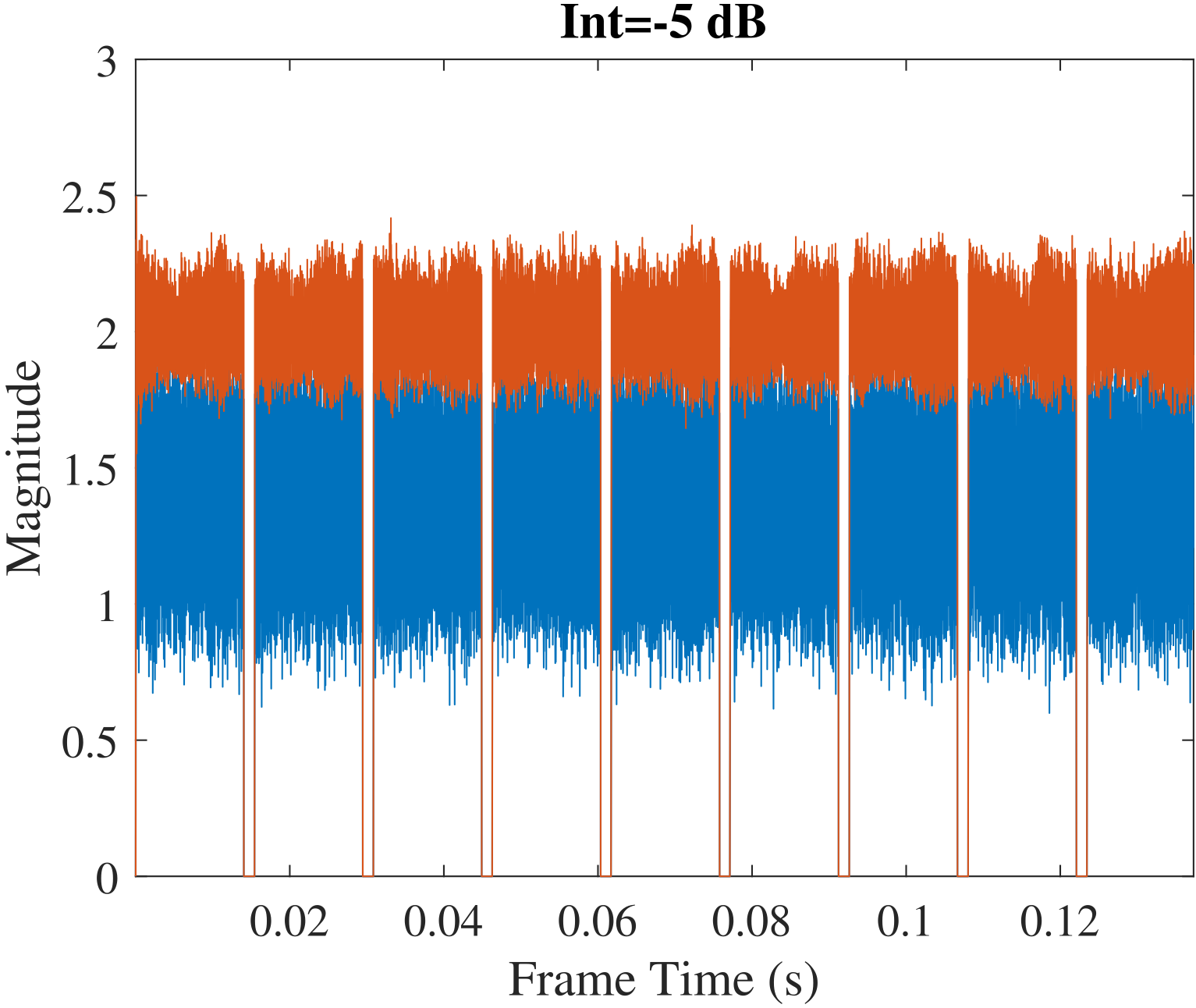}}\\
   \subfloat[]{\includegraphics[width=0.4\columnwidth]{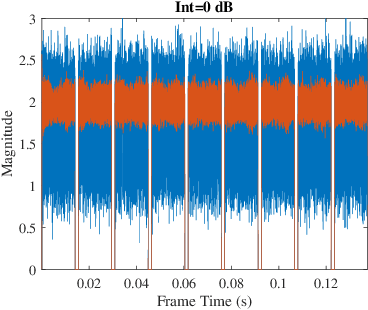}}\\
    \caption{Received waveforms as shown in blue and optimized waveforms as shown in orange for various interference levels (a) -10 dB (b) -5 dB (c) 0 dB.}
    \label{fig:bob_Ints}
\end{figure}

The experiment begins with a degraded channel and performs RIS-assisted optimization of the link. USRP I was set to operate at sampling rate of 400 kS/s and the USRP configuration parameters are summarized in Table \ref{tab:table1}. 
The QPSK modulated signals transmitted in the main communication link were generated using Simulink according to previous work which details the Simulink baseband models  \cite{lodro2020near}\cite{lodro2020mimo}. 
The base-band interference signals were generated using GNU Radio and transmitted using USRP II at 5.2 GHz. 
Typical waveform sent by Alice is shown in Fig.\ref{fig:alice} and the typical noise waveform sent by Eve is shown in Fig. \ref{fig:eve}. Furthermore, typical received and optimized waveforms at Bob are shown in Fig. \ref{fig:bob_Ints} with relative time frame. The waveforms shown at Bob are at different levels of interferences.

We avoided any packet loss risk in the communication between the SDR and host PC over Ethernet cable by performing IQ data acquisition at lower sampling rates. Packet loss can be observed by monitoring the presence and absence of long string of ones and zeros at the terminal.
Furthermore, we monitored the transmitter and receiver system object, as the presence of overflows may lead to inconsistent performance. 
USRP I and II were set to stream data in real-time, hence reliable EVM measurements were recorded that could only be influenced by the propagation environment in the cavity. 
The EVM was regularly measured in a time window of 10 s from an stable QPSK constellation diagram after all the hardware impairments from the received waveform were removed.

\begin{table}[]
    \centering
    \begin{tabular}{|c|c|}
        \hline 
        Parameters & Values \\
        \hline 
        Sampling Rate & 400 kS/s\\
        Master Clock Rate & 200 MHz\\
        Interpolation/Decimation & 500\\
        Tx Channel & Channel 1\\
        Rx Channel & Channel 2\\
        Tx Gain & 1 dB\\
        Rx Gain & 0 dB\\
        \hline 
    \end{tabular}
    \caption{Main communication link USRP parameters}
    \label{tab:table1}
\end{table}
\section{RESULTS}

\subsection{RIS-assisted channel hardening: NLOS without intentional interference}
\begin{figure}
    \centering
	\includegraphics[width=0.6\columnwidth]{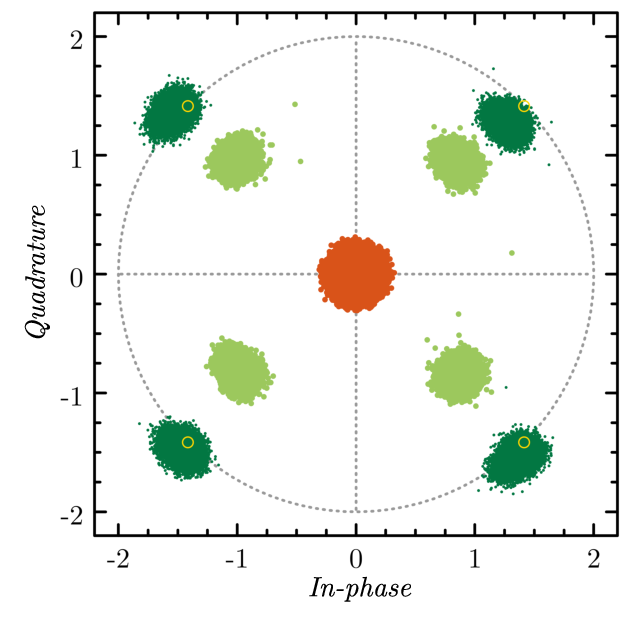}
	\caption{Constellation diagrams between Alice and Bob  for three different scenario. In red, both RIS are switched off. In light green, only one RIS is used to optimize the mean EVM. In dark green, both RIS are used to optimize the mean EVM. The gold circle correspond to ideal symbols for QPSK modulation. In all scenario  EVE is switched off. \label{fig:fig2}}
\end{figure}
In presence of environment related interference generated from multi-path 
propagation, the wireless link between Alice and Bob operates unperturbed as Eve does not interfere with QPSK symbols. 
Starting from a random configuration for both the two RISs, we measure the constellation diagram, red cloud of points in Fig. \ref{fig:fig2}, and in each quadrant of the complex plane, we evaluate 
\begin{equation}
\sigma_{\mathrm{k}} = \frac{\left | \sigma_\mathrm{k}-\sigma_\mathrm{I}\right |}{\sigma_\mathrm{I}} \label{EVM}
\end{equation}
where $\sigma_\mathrm{k}$ is the error vector magnitude between the $\mathrm{k}$-th actual symbol sent associated to the complex value  $s_\mathrm{k}$ and the ideal QPSK symbol $s_\mathrm{I}$, gold circles in Figure~\ref{fig:fig2}. 
Then, we compute the mean EVM, hereafter denoted by $\sigma_\mathrm{QPSK}$, and defined as the root mean square of the ensemble of $\left\lbrace \sigma_\mathrm{k}\right\rbrace$
\begin{equation}
\mathrm{EVM}=\sqrt{\frac{1}{N}\sum_{\mathrm{k}}^{N} \sigma_\mathrm{k}^2}.
\end{equation}
We are interested in the mean EVM  as it is a KPI that probes the quality of the digital communication link, and it is directly obtained using EVM block in the baseband Simulink model. 
In other words, we do not need to post-process data externally and the EVM is measured in real-time in the baseband model. 
The red cloud shown in  Fig. \ref{fig:fig2} corresponds to the  constellation diagram measured with the initial random configuration of the RIS, and its  mean EVM is $\mathrm{EVM}=90\,\%$. 
Starting from this initial situation, we then use the RIS to improve the communication link between Alice and Bob by minimizing the mean EVM.  
To do so, we use the iterative optimisation algorithm described in \cite{Mosk2012}. 
The state of the first pixel is flipped to its opposite state and kept in this new state, if the goal function (here the  average EVM) decreased. 
Otherwise, the pixel is flipped again back to its previous state.
This procedure is repeated with all pixels one by one and typically requires multiple loops across the RIS reflection matrix.
The optimization is terminated when the goal function no longer decrease during one loop of the optimization.
Because it based on root mean square, the minimization of the mean EVM simultaneously allows us to reduced the distance between the barycentres of the measured constellation and the ideal symbols (gold circle in Figure~\ref{fig:fig2})
and the spreading of the clouds of points around those barycentres.
It is worth noticing that we are optimizing measured quantities that are related to the performance of the \emph{digital} wireless data link, instead of the SNR. 
In particular, we obtain the simultaneous increase of both the average and the variance of symbol energy by minimizing the average EVM.
For this  experiment we first optimize the configuration of only one RIS, keeping the state of the second one unchanged. 
The next step, starting from the final optimized configuration, is to include both RISs in the optimization process. 
The final constellation  resulting from the optimization involving one or two RIS  are shown in Fig. \ref{fig:fig2}. The light green clouds of points correspond to the configuration where only one RIS is used. The mean EVM in this case moves from  $\mathrm{EVM}=90\,\%$ to $\mathrm{EVM}=36.4\,\%$.  The dark green clouds is the constellation  diagram   resulting from  the optimization with both RIS. In this case, the mean EVM is $\mathrm{EVM}=8\,\%$ and the measured symbols are really close to the ideal symbols.  In both scenarios, when we use the RIS we are able to improve the communication link between Alice and Bob by optimizing in real-time the mean  EVM given by the SDR set-up. 
Of course, the larger the controllable surfaces the higher the degrees of freedom to control the EM field inside the chamber, and the closer we can get to ideal constellation symbols. 
Indeed, it is known from \cite{Dupre2015}, that the amount of power we are able to focus with RIS is directly linked to the number of controllable degrees of freedom. 
Therefore, in comparison with the scenario using two RIS elements, against only one RIS in the present experiment, we achieve enough wave field control to separate the symbols. 
However, the power that we are able to focus on Bob is limited, and the distance between cloud barycenters and ideal symbols is bounded. 
This is mainly due to the limited number of RIS pixels, as well as to the presence of short orbits \cite{Anlage2010} that do not intercept the RIS, and thus couple transmitter and receiver through the multi-path propagation environment. 
Moreover, the width of the cloud associated to the constellation symbols after RIS iterative optimization is also bounded.  
Those fundamental limitations to the RIS operation within a reverberating environment are mainly ascribed to: i) the finite loss factor of the cavity, established by imperfectly conducting walls; ii) the finite RIS area covering the cavity walls, which limits spatial self-averaging underpinning the channel hardening effect. 
Both the effects contribute to form the residual fluctuation of the constellation symbols, in addition to thermal noise fluctuations at receiver.
In particular, based on \cite{Dupre2015} and the fluctuation of the quality factor of overmoded enclosures \cite{Arnaut2012}, we expect that the radius of the fluctuation cloud is proportional to the standard deviation of the cavity quality factor. Hence, $\sigma_Q \propto \sqrt{1/M}$ where $M$ is the maximum number of ergodic cavity eigenmodes controlled by the RIS in the optimal configuration. 
The maximum data-rate is subject to the same limitations of the average SNR achievable by a prescribed number of RIS pixels. 
However, the CH effect accompanying optimal RISs makes the wireless communication link robust against multipath fading and co-channel interference, as explained in the next subsection, even with a moderate number of controlled ergodic eigenmodes. 
This has been recently ascribed to the law of large numbers for RISs with an increasing number of independent pixels \cite{Sanguinetti2021}. 

\subsection{RIS-assisted channel hardening: NLOS with intentional interference} 
\begin{figure}
    \centering
	\includegraphics[width=0.5\columnwidth]{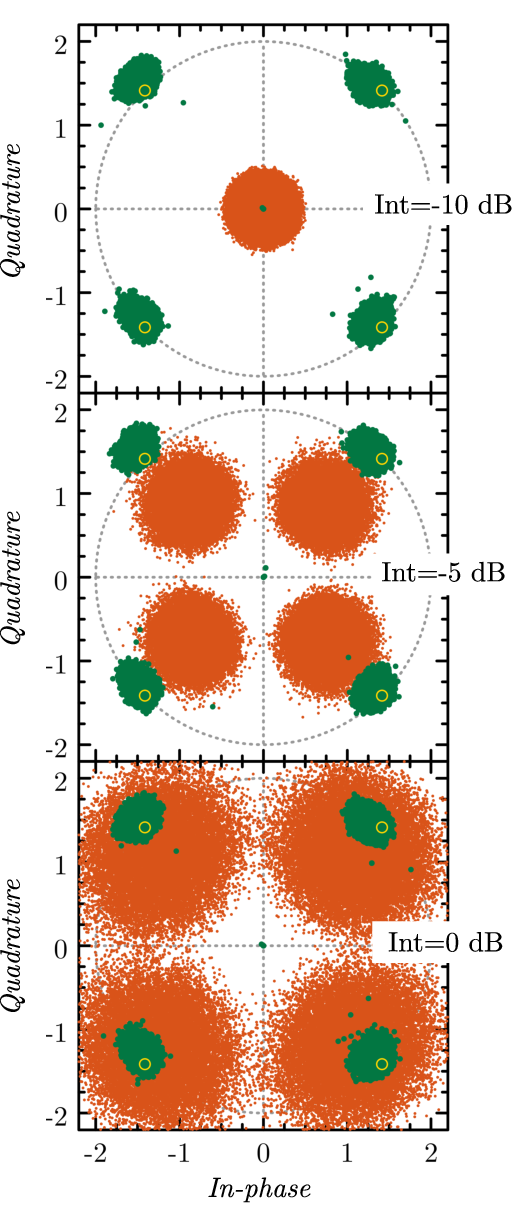}
	\caption{ Constellation diagrams between Alice and Bob when EVE is switched on and it is acting as a source of co-channel interference. From top panel to bottom panel, the intensity of interferences is sequentially increased by step of $5$~dB, starting from interferences with an intensity of $10$~dB lower than the signal between Alice and Bob. The red constellations correspond with the constellation at the time of activation of EVE. The dark green constellations are the constellation obtained after re-optimization of the mean EVM using both RIS. The gold circle correspond to ideal symbols for QPSK modulation.  \label{fig:fig3}}
\end{figure}
In this section, we report the ability of RIS to cancel unwanted co-channel interference caused, for instance, by a jammer or simply other emitters in the environment perturbing the channel. 
To do this, we use as the sources of interference the signal provided by USRP II (Eve)in Figure~\ref{fig:fig1}.  
When Eve is switched on, the transmission between Alice and Bob is attacked by the co-channel interference presented by Eve who emitted in the environment Gaussian noise with tunable intensity. 
The waveform transmitted by Eve can be seen in Fig. \ref{fig:eve} which was transmitted for three levels of unintended interference. 
The ratio between the signal between Alice and Bob and Eve interference is defined as $\textrm{Int}$. 
Starting from a random RIS configuration and $\textrm{Int}=10$ dB, we first optimize the mean EVM using both RIS, then when the optimisation reaches a plateau, we increase the interference intensity by $5$~dB and re-optimized the mean EVM until a new plateau is reached. 
This procedure is repeated until the signal and interference strength are equal. 
The corresponding constellation diagrams are shown in Fig.\ref{fig:fig3}. 
From top to bottom panel, the interference intensity is increased at steps of $5$~dB. 
The red constellations on each correspond to the constellation measured at the time when Eve power is increased. 
From top to the bottom , the mean EVM associated to these red constellations are
\begin{itemize}
	\item $\mathrm{EVM}=90\,\%$ for interference strength  $10$~dB lower than the signal strength and random configuration of RIS
	\item  $\mathrm{EVM}=45\,\%$ after first optimization and an interference to signal power ratio  $\textrm{Int}=-5$~dB 
	\item  $\mathrm{EVM}=37\,\%$ after the second optimization and  an interference to signal power ratio $\textrm{Int}=0$~dB. 
\end{itemize} 
Let us note that the more the SINR increase the more the constellation symbols are spread through the complex plane. 
In each panel of  Fig.\ref{fig:fig3}, dark green constellations are the ones obtained after the optimizations that follow the increase of interferences' power.  
Whatever the interference to signal power ratio, the optimized constellation barycenter returns to be close to ideal symbols, with a mean EVM  $\mathrm{EVM}=8\,\%$. 
We thus show the ability of RIS to mitigate and even cancel the co-channel interference operated by Eve. 
Indeed, we would like to draw the attention of the reader to the fact that all optimized constellation diagrams shown in Figure~\ref{fig:fig3} are similar to the one shown in Figure~\ref{fig:fig2}, which are associated to the case with no interference and both RISs used for the optimization.

\section{Conclusion}
We have shown that robust multi-path fading mitigation can be achieved by using reconfigurable intelligent surfaces within an overmoded  metallic enclosure supporting rich multipath fading. 
The adopted environment mimics complex wave propagation in wireless indoor channels, while controlling distributed losses within its interior volume.
We have found that RIS optimization results in channel hardening, besides energy maximization widely reported in the scientific literature. 
A software defined radio system makes it possible to observe channel hardening directly in the constellation of a QPSK digital modulation. 
Achievements show the RIS capability of optimizing wireless links under combined multi-path fading and intentional/unintentional co-channel interference, making it an appealing technology for operation in contested electromagnetic environments such as multi-operator mobile wireless networks. 

\section{Acknowledgment}
The work of GG was supported in part by the Royal Society under Grant INF\textbackslash R2\textbackslash 192066; in part by the EPSRC under Grant EP/V048937/1, and in part by the European Commission through the H2020 RISE-6G Project under Grant 101017011. Greenerwave acknowledges funding from the French “Ministere des Armées, Direction G ́ne ́rale de l’Armement” and “Agence de l’Innovation de de ́fense” through the RAPID “3SFA” project and the RAPID “M3SFA” project.


\appendix

\section{USRP Architecture and performance}
\begin{figure}
    \centering
    \includegraphics[width=0.8\columnwidth]{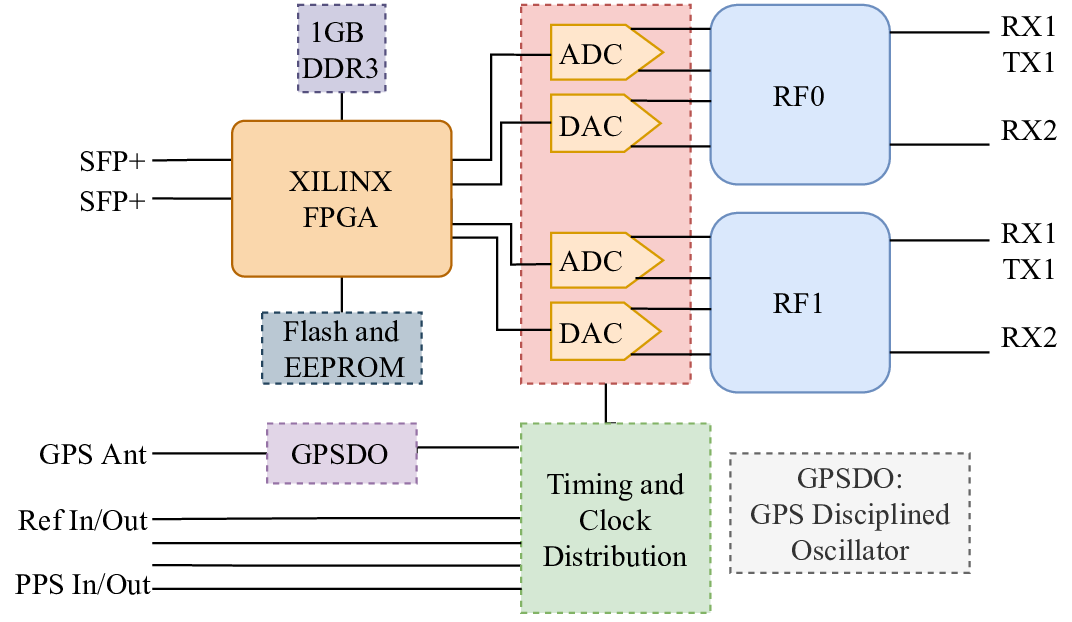}
    \caption{X310 USRP architecture}
    \label{fig:x310}
\end{figure}
Figure \ref{fig:x310} shows an architecture of USRP I. 
The USRP I has: i) modular structure where RF cards can be separately interfaced with the motherboard; ii) high performance FPGA which is interfaced with RF frontends, host interfaces, and DDR3 memory. 
When used with UBX-160 cards, the USRP I can provide a maximum bandwidth of up to 160 MHz, and it is tunable from 10 MHz to 6 GHz. 
It has a high-speed 14-bit ADC and 16-bit DACs per RF chains with maximum clock rates of 200 MS/s and 800 MS/s respectively. Additionally, USRP I has a clock and timing distribution circuitry to distribute external 10 MHz and 1 PPS synchronization signals, 
and has out-of-the-box support for 1 GigE and 10GigE that can be harnessed using SFP+ adapters. 
However, the exploitation of maximum sampling rates depends upon host capabilities, including waveform complexity, host PC speed, and writing speed of memory.

\end{document}